\documentclass{article}

\usepackage{arxiv}

\usepackage{graphicx}
\usepackage{dcolumn}
\usepackage{amsmath,amsfonts,amssymb}  
\usepackage{bm}
\usepackage{color}
\usepackage{hyperref}
\usepackage{colortbl,xcolor}
\definecolor{green}{cmyk}{0.75002 0 1 0}
\definecolor{yellow}{cmyk}{0.04 0.3 1 0.02}
\definecolor{orange}{cmyk}{0 0.6 1 0}
\definecolor{blue}{rgb}{ 0    0.3470    0.7410}
\definecolor{red}{rgb}{1 0.2 0}

\title{Extreme Aerodynamics: A Data-Driven Perspective}

\author{
  Kunihiko Taira\\
  Department of Mechanical and Aerospace Engineering\\
  University of California, Los Angeles, California 90095, USA\\
  \texttt{ktaira@seas.ucla.edu} 
}

\date{November 16, 2025}

\begin{document}

\maketitle

\begin{abstract}
While experiencing atmospheric turbulence on a commercial flight can be uncomfortable, it rarely compromises the stability of the aircraft.  The situation is quite different for small air vehicles that operate in urban canyons, around mountainous terrains, and in the wakes of marine vessels, where they could encounter highly unsteady atmospheric conditions with relatively strong gusts.  The spatiotemporal scales of such disturbances can be larger than the characteristic aerodynamic scales of the small vehicles, making the relative effect of disturbance significantly stronger than what a large commercial aircraft would experience.  The gust ratio can exceed 1 in these extreme flight environments, making stable flight difficult, if not currently impossible.  We refer to the study of aerodynamics for gust ratios over 1, {\it extreme aerodynamics}, and identify major challenges that require breakthroughs, particularly with data-driven approaches.  Extreme aerodynamics present unique opportunities for innovative analysis techniques to study rich flow physics problems with strong nonlinearity, transient dynamics, and low-dimensional modeling over a large parameter space.  Some of the approaches discussed herein should apply to a wider range of fluid dynamics problems with similar challenges.

\end{abstract}


\section{Introduction}

In-flight encounters with atmospheric turbulence are uncomfortable.  In some instances, turbulence can cause your favorite beverage to spill and ruin your clothes or, even worse, damage your laptop computer on which you are furiously preparing your upcoming presentation.  When the level of turbulence is high, the unsteadiness may cause unfortunate injuries in flight, which commercial airlines strive to avoid by flying around such adverse airspace when possible \cite{deMedeiros:JAS25}.  Even if these unpleasant encounters with adverse weather events may seem unnerving, modern commercial airliners are designed with high structural integrity and flight control algorithms to safeguard the aircraft.

The situation can be drastically different for smaller aircraft that operate at lower altitudes with lower speeds.  Operating smaller aircraft such as drones and personal air vehicles in calm weather on a sunny day is, of course, not an issue.  However, flight conditions in adverse weather with strong winds can be particularly challenging over mountainous terrains, in urban canyons, and around large-scale structures.  In those airspaces, a multitude of vortical structures in fluctuating velocity fields can impact the flying vehicles \cite{Jones:ARFM22,Jones:AIAAJ21}.   These sudden changes in the air appear as enormous transients in freestream velocity and a sea of strong vortices with characteristic scales comparable to the scales of the wing or the aircraft.  These disturbances can render air vehicles unable to maintain stable flight \cite{Jones:ARFM22,BajajIEEE20,Gao:SR21}. 

For these reasons, smaller vehicles are forced to stay on the ground or avoid such violent airspace altogether when the relative unsteadiness is significantly large.  Even with these challenges, there are instances in which critical rescue or defense missions must be carried out.  Moreover, there may be situations in which a small-scale aircraft, such as a personalized air vehicle or air taxi, may unexpectedly encounter an extreme flight environment without warning due to the sharp changes in microweather, which appears to be occurring at a greater frequency \cite{Dim:NCC12}.  Having the technical capability to maneuver an air vehicle in such an environment is necessary.  

Conventional aerodynamic theories and technologies are founded on the assumption of aircraft being in steady or quasisteady flight conditions.  Most aerodynamic studies are based on linear analysis about a time-invariable operating condition with nonlinear extensions for small levels of perturbations \cite{wagner1924entstehung,glauert1930force,theodorsen1935general,Sears:JAS41,Hoblit88,Katz01,Leishman06}.  In contrast, there is no established aerodynamic theory for extremely unsteady aerodynamic flows, in which atmospheric disturbances with large amplitudes and fluctuations are on the time scale commensurate with those of the aircraft wake.  Over the last decade, there have been a number of experimental, computational, and theoretical studies focused on the influence of discrete gusts on wings \cite{Jones:AIAAJ21,Jones:ARFM22}.  Most of these recent efforts have studied modest levels of disturbances, and their collective work has amassed a valuable number of datasets aimed to enhance the fundamental understanding of gust-wing interactions \cite{Barnes:AST18,Medina:AIAA20,Biler:AIAAJ21,Mohamed:Drones23,Chen:AIAAJ24,Li:AIAA25}.

Here, we consider the effect of strong gusts on the aerodynamics of wings.  We particularly focus on gusts with characteristic velocities larger than the flight speed (or freestream velocity).  The ratio between the gust velocity $u_G$ and the freestream velocity $U_\infty$ is defined as the gust ratio $G \equiv u_G / U_\infty$ and is one of the important parameters that quantifies the strength of the gust.  In this paper, we refer to aerodynamics with $G > 1$ as {\it extreme aerodynamics}.  When $G > 1$, the interaction between the wing and the disturbance is highly nonlinear, with violent vortex dynamics taking place around the wing.

Turbulent disturbances in the atmosphere can be of various forms (e.g., disturbance size, strength, position, orientation, and Reynolds number) \cite{Jones:PRF20,Stutz:AST23}.  If these disturbance parameters are naively considered, the investigation would require an astronomical number of parametric studies to be performed experimentally and computationally.  To this end, the study of extreme aerodynamics requires a shift in our scientific perspective from traditional aerodynamics to a study that assimilates an enormous collection of cases and extracts the dominant influence of strong disturbances on flying bodies.

This major problem translates to the identification of universal nonlinear dynamics from an enormous amount of data in an effective and efficient manner while gaining physical insights into extreme aerodynamics.  That is, there is a need to reveal the underlying physics that captures the primary extreme aerodynamic response over an appropriate set of feature variables.  Data-driven techniques \cite{Brunton:ARFM20,Brunton_Kutz_2022,Bishop2006,Goodfellow-et-al-2016,Sutton:RL,Watt20,Bishop24,Taira:PRF25} can serve as a powerful tool to analyze the vast data set and reveal the underlying dynamics in seemingly complex flow phenomena.  Here, we remind ourselves that the proposed effort is not a mere machine learning exercise.  There needs to be significant data reduction to capture the flow physics over the right set of variables (coordinates) while being informed by physics, and to ensure the gained insights are interpretable and beneficial for future aircraft operations and designs.

The purpose of this article is to call attention to the challenges in the study of extreme aerodynamics.  In particular, we discuss the need for developing a novel and innovative framework to detect the approach of unknown large-scale perturbations in the atmosphere, determine how such disturbances affect the aerodynamics of the vehicle, and perform flow control or maneuvers to achieve stable flight.  There are also a number of other issues that make flying in extreme environments challenging.  The examples presented herein are based on the activities undertaken by the author's research group and are generally based on datasets collected from numerical simulations.  The perspective below is provided in hopes of stimulating exchange of ideas on novel data-driven analysis techniques together with experimental, computational, and theoretical studies \cite{Taira:PRF25} for not only extreme aerodynamic problems but also for those that involve strong disturbances and transient physics.  At the end of this paper, we provide a brief discussion on the outlook of using data-driven techniques for extreme aerodynamics.


\section{Atmospheric Disturbances}

Atmospheric disturbances are comprised of localized vortices, jets, and shear layers of various sizes, strengths, and orientations \cite{Fernando:BAMS19,Jones:ARFM22}.  Gust models aim to capture the effect of these discrete and continuous disturbances.  
For traditional large-scale high-speed aircraft, freestream perturbations are considered to be modest in their amplitudes over long spatial or temporal scales \cite{Hoblit88,Leishman06}.  The effects of these perturbations on larger aircraft are generally felt over many chord lengths of travel.

When the spatial and temporal scales associated with the atmospheric disturbances are comparable to those of small-scale aircraft, there can be a high level of aerodynamic forces and moments exerted on the aircraft in a transient manner over a short period of time.  Moreover, when the amplitude of the gust velocity is comparable to or larger than the flight speed, the nonlinear aerodynamic effects become significant with the possibility of gross modification to the flow around the vehicle, including massive separation.  

To gain a deep understanding of the impact of these strong gust disturbances on the wing aerodynamics, it is important to select appropriate gust models that capture the essence of the violent disturbances.  
Over the last decade, the effects of discrete gusts on wing aerodynamics have been extensively studied \cite{Jones:AIAAJ21,Jones:ARFM22}.  While most of these recent studies handle small to modest gust ratios $G \lesssim 1$, they provide great insights into the nature of unsteady aerodynamic responses to gusts and the vortex dynamics that evolve around the wings.  The present research efforts build upon these prior investigations and consider discrete gusts in the form of vortices and jets.

For continuous gusts, we require insights from real flight environments, such as those in the vicinity of mountains, canyons, buildings, and ships \cite{Fernando:BAMS15,Fernando:BAMS19,Eliasson:AE06,Zajic:JAMC11,Yang:BE23,Mohamed:Drones23,Ilyas:PF25,Hu:JFM23,Cao:BE25,Forrest:CF10,Thedin:JA20,Dooley:CF20}.  In many of these studies, the spatial and temporal resolution of the atmospheric measurements are usually not sufficiently fine to fully capture the richness of the gusty conditions.  Meteorological research collects weather data in some of these environments, but their requirements on spatial and temporal scales tend to be much coarser than what aerodynamic research necessitates to achieve situational awareness during flight.  As such, there may be a need to take coarse experimental measurements and numerical simulations and reconstruct the flow accurately to serve as a continuous gust model.  This type of effort invites the use of machine-learning-based super-resolution analysis \cite{Fukami:JFM19,Fukami:JFM21,Fukami:TCFD23} to provide micro-scale weather estimation and forecast based on sparse sensors.  This may be achieved in an urban area where some sensors are pre-installed.

With the high-resolution flow fields being available from simulations or experiments, important flow features can be extracted to derive gust models useful for extreme aerodynamic studies.  In particular, scale decomposition analysis \cite{Goto:PRF17,Fujino:JFM23} may help identify critical flow structures with scales comparable to the flying body.  Use of such a decomposition technique can also be combined with force element/partitioning analysis \cite{Chang:PRSLA92,Lee:JFM12,Zhang:JFM20,Menon:PRF22} to uncover which structures are dangerous.  There are also works based on information theory \cite{Lozano:PRR22,Arranz:JFM24,Fukami:AIAAJXX} that can decompose the flow into informative and non-informative structures, which may reduce the number the gust disturbance structures that need to be tracked and modeled.  By extracting structures that are important to extreme aerodynamic flight through these data-driven techniques, we should be able to develop low-order continuous gust models that accurately replicate the dominant effects of extreme gusts on air vehicles.

\section{Extreme Aerodynamic Flows}
\label{sec:vtx_impinge}

\begin{figure}
	\centering
	\includegraphics[width=0.99\textwidth]{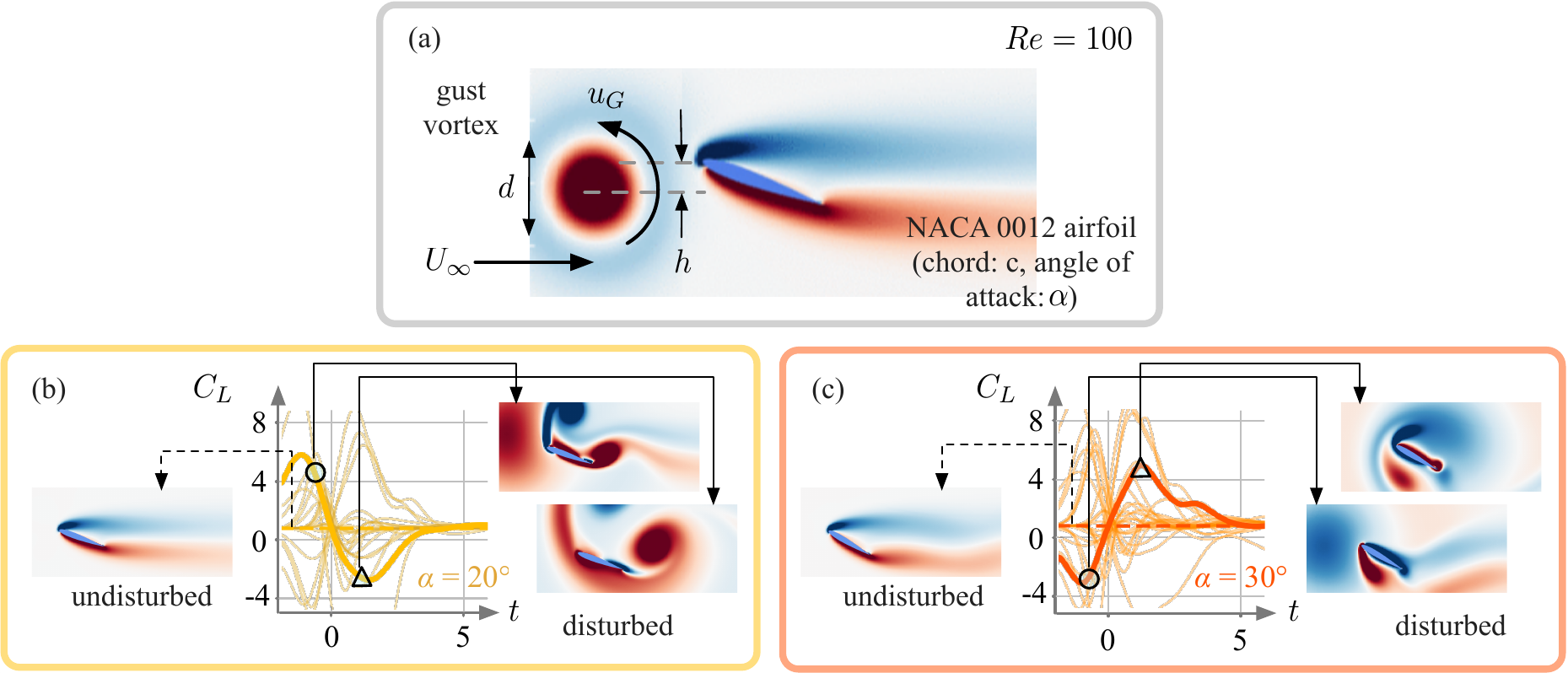}
	\caption{Extreme aerodynamic flows with gust vortex impinging on a NACA 0012 airfoil at $Re = 100$.  (a) Problem setup.  A collection of lift traces and representative vorticity fields shown for (b) $\alpha = 20^\circ$ and (c) $\alpha = 30^\circ$.  
    Figures adapted from \cite{Fukami:NC23}.}
    \label{fig:vtx_imp}
\end{figure}

Let us present the complexity of extreme aerodynamic flows through an example of a strong discrete gust vortex impinging on an airfoil \cite{Fukami:NC23,Fukami:JFM24}.  In this example, we consider a two-dimensional laminar flow with a NACA 0012 airfoil at a chord-based Reynolds number of $Re \equiv U_\infty c / \nu = 100$, where $c$ is the chord length and $\nu$ is the kinematic viscosity.  Here, the gust is modeled as a Taylor vortex with a maximum angular velocity of $u_G$ at radius $d/2$, initially positioned at a vertical distance $h$ away from the leading edge, as presented in figure \ref{fig:vtx_imp}(a).  The non-dimensional parameters that are varied in this example are the gust ratio $G \equiv u_G/U_\infty$, size $d/c$, and vertical position $h/c$, as well as the angle of attack $\alpha$ of the airfoil.

Even though this setup may appear simple, the problem quickly becomes challenging in terms of tractability.  If $m$ discrete choices are made for each of the parameters ($G$, $d/c$, $h/c$, $\alpha$), there will be $m^4$ cases to be studied.  Furthermore, for a flow requiring spatial discretization of $n_x \times n_y$ points with $n_t$ temporal snapshots, we will be collecting a total of $m^4 n_x n_y n_t$ elements.  Based on conservative estimates of $m = 5$, $n_x = n_y = 100$, and $n_t = 1,000$, the overall study for the laminar extreme aerodynamic problem will necessitate an examination of $m^4 n_x n_y n_t = 6.25 \times 10^9$ elements of data.  For other problems, the number of parameters and grid points can be substantially larger, making the overall number of elements to be stored and examined astronomical if the parametric study is not carefully designed (especially for three-dimensional turbulent flow cases).  As evident from these estimates, gust response analysis requires a systematic examination of this large flow field dataset.  Appropriate sampling and flow field analysis calls for innovative machine learning approaches to efficiently interpret the nonlinear flow physics \cite{Montgomery19,Fuhg:ACME21,Vinuesa:NRP23,Sotoudeh:AIAAJ23}.

Now, let us turn our attention to the unsteady flows generated by the gust vortex impingement.  For example, consider a case with ($G$, $d/c$, $h/c$, $\alpha$) = ($3.8$, $2$, $0.1$, $20^\circ$) shown in figure \ref{fig:vtx_imp}(b).  Here, a strong vortex with positive circulation hits the airfoil at $\alpha = 20^\circ$ with an initially steady wake.  When the gust vortex (red) impinges on the airfoil, a strong negative vorticity sheet is generated from the leading edge in response.  This sheet rolls up into a leading-edge vortex \cite{Eldredge:ARFM19}, exerting a massive transient increase in lift (thick lift curve).  Concurrently, the flow near the trailing edge is also modified, resulting in the creation of a trailing-edge vortex (red).  As the leading-edge vortex detaches from the airfoil, the trailing-edge vortex continues to grow over time and exerts a large negative lift force.  During this extreme vortex gust encounter, the airfoil experiences massive separation with the creation of two large vortices.

The above description is only for one case of extreme aerodynamic flow.  Another case is shown for $\alpha = 30^\circ$ in figure \ref{fig:vtx_imp}(c).  Here, the unperturbed wake exhibits periodic wake oscillation due to the Karman shedding.  In the highlighted case, a negative sign vortex (blue) hits the wing and exhibits a reversed trend in exerting transient lift fluctuations.  While the airfoil also encounters massive separation with very large vortical structures generated around itself, the behavior of these structures is quite different from the previous case.  Generally, for a gust ratio of $G > 1$, the lift can fluctuate tenfold over a couple of convective time units.  The exact temporal profile of the force fluctuation can vary depending on the parameters ($G$, $d/c$, $h/c$, $\alpha$), as shown in figures \ref{fig:vtx_imp}(b) and (c).  In addition to the discussed cases, figures \ref{fig:vtx_imp}(b) and (c) present lift histories from a number of other cases with curves in a lighter shade.  As it can be seen, a simple collapse of data cannot be easily achieved, given the rich flow physics and the vast coverage of the parameter space.  While this particular flow setup considered $Re = 100$, analogous flows for a spanwise periodic airfoil at $Re = 5,000$ also share qualitatively similar flow characteristics even in the presence of fine-scale flow structures that emerge from instabilities \cite{Fukami:PRF25}.  This point is further discussed from the perspective of vortex dynamics in Section \ref{sec:vortex_dynamics} and illustrated with an analogous three-dimensional extreme aerodynamic example \cite{Odaka:JFMXX} in figure \ref{fig:AR1}.

In extreme aerodynamic flows, it is important not only to consider (nominally) steady unperturbed flows at low $\alpha$ as initial conditions but also to consider unsteady initial conditions at high $\alpha$.  Because extreme gust encounters impose massive separation and with large-amplitude transient behavior, it is important to incorporate as much unsteadiness as possible, even from the beginning, to include rich wake response in the collected training data.  This becomes a very important point for machine learning efforts to ensure that the training dataset holds rich physical information on extreme aerodynamics as much as possible.  Furthermore, gradually increasing the complexity of the flow can be beneficial in promoting the understanding of the physics and in implementing data-driven techniques successfully.  For example, transfer learning \cite{Weiss:JBD16} can be used to analyze and model flows at higher Reynolds numbers based on the lower Reynolds number results, without the need to redo the training process \cite{Inubushi:PRE20}.

\section{Extreme aerodynamic attractors}
\label{sec:manifolds}

When studying the vast response of extreme aerodynamic flows to a multitude of gust parameters and types, a giant collection of flow field data from numerical simulations and experiments is accumulated.  Across the collected cases, it is desirable to find the universal flow physics.  Due to the large number of cases, manually studying each and every case and deducing similarities is difficult, if not humanly impossible.  For this reason, a systematic approach is necessary to extract the underlying dominant physics.  One such approach could be modal analysis \cite{Taira_etal:AIAAJ17,Taira_etal:AIAAJ20}, which can identify important spatial modes that capture the dominant flow features.  Proper orthogonal decomposition \cite{Berkooz:ARFM93,Sirovich:QAM87} can help with compressing the flow field snapshots into key features.  However, due to the linear nature of the compression, it is challenging to compress the flow field information to an interpretable low-dimensional subspace (additional discussions on this matter are provided in Section \ref{sec:vortex_dynamics} and \cite{Linot:FDR25}).   Moreover, due to the linear nature of the compression, the reconstructed flow from the compressed state can exhibit artifacts of the data collection setup \cite{Fukami:NC23}.  

\begin{figure}
	\centering
	\includegraphics[width=0.7\textwidth]{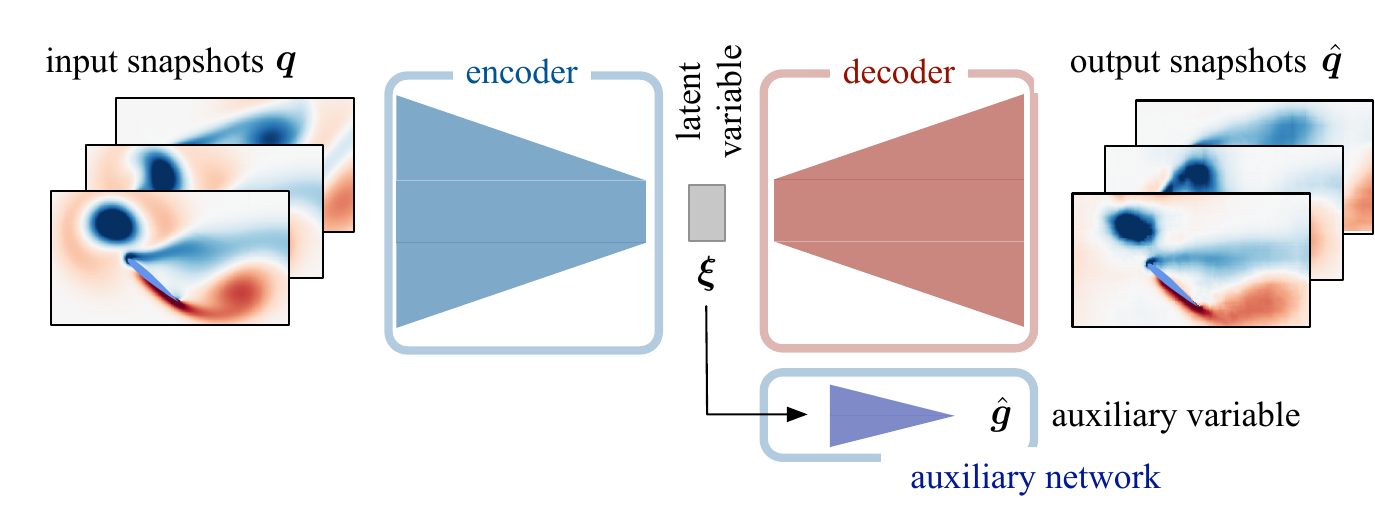}
	\caption{Compression of extreme aerodynamic flow snapshots using a convolutional neural network-based autoencoder \cite{Fukami:NC23}.  This network features observable augmentation using an auxiliary network on the side.}  
    \label{fig:AE}
\end{figure}

Alternatively, we can take advantage of nonlinear data-driven compression techniques such as an autoencoder \cite{Hinton:Science06,Fukami:NC23,Ozalp:ND25} to find a low-dimensional latent space representation of the extreme aerodynamic flows.   With this machine-learning-based compression of the large flow field dataset, recent studies have found that the transient flows from the aforementioned dataset can be compressed to a low-dimensional latent variable \cite{Omata:AIP19,Linot:PRE20,Fukami:NC23,DeJesus:PRF23,Smith:JFM24,Odaka:AIAA25,fukagata2025compressing,Ozalp:ND25}.  To achieve compression of the vast flow field data, let us consider an autoencoder shown in figure \ref{fig:AE}, which is comprised of two main components.  First, the encoder $\mathcal{F}_E$ compresses the input flow field $\boldsymbol{q}$ to a low-dimensional latent variable $\boldsymbol{\xi}$.  Second, the decoder $\mathcal{F}_D$ decompresses $\xi$ to the output flow field that is supposed to be a reproduction $\hat{\boldsymbol{q}}$ of the input variable.  This process can be expressed as
\begin{equation}
    \hat{\boldsymbol{q}} = \mathcal{F}_D (\boldsymbol{\xi}) = \mathcal{F}_D ( \mathcal{F}_E ( \boldsymbol{q} ) ).
\end{equation}
In addition to these two main components, an auxiliary network $\mathcal{F}_A$ can be added to the side to reproduce an observable $\boldsymbol{g}$ (e.g., forces on a body or sensor measurements) \cite{Fukami:NC23}
\begin{equation}
    \hat{\boldsymbol{g}} = \mathcal{F}_A(\boldsymbol{\xi}).
\end{equation}
In this formulation, the weights $\boldsymbol{\theta}$ in the autoencoder network are found by solving 
\begin{equation}
    \boldsymbol{\theta}^* = \underset{\boldsymbol{\theta}}{\mathrm{argmin}} 
    \left( 
        \| \boldsymbol{q} - \hat{\boldsymbol{q}} \| + \beta \| \boldsymbol{g} - \hat{\boldsymbol{g}} \|
    \right).
\end{equation}
When the reproduced variable $\hat{\boldsymbol{q}}$ is close to ${\boldsymbol{q}}$ based on a low-dimensional $\boldsymbol{\xi}$, compression is considered successful and $\boldsymbol{\xi}$ is regarded to hold the minimally required information to capture the variations in $\boldsymbol{q}$.  Here, the norm can be selected appropriately to define the loss function (e.g., $L_2$ or weighted).  The size of $\boldsymbol{\xi}$ is chosen such that the error reaches some satisfactory level while being as small as possible in its dimension.  Details on this type of analysis can be found in \cite{Fukami:NC23}.

The addition of the side network is beneficial in providing physical importance to $\boldsymbol{\xi}$ and its interpretability.  We also note that the use of a convolutional neural network \cite{LeCun1998} within the autoencoder formulation is important for analyzing field variables such as the velocity or vorticity field, for which the spatial arrangement of the physical variables is important.  Convolutional neural networks use a filtering process to identify certain spatial features from the field variable.  This is critical in analyzing extreme aerodynamics, where various flow structures emerge over a large spatial domain.  Alternatively, vision transformers \cite{dosovitskiy2020image} could also support the identification of flow structures.

Now, let us revisit the problem of extreme gust vortex impinging on an airfoil presented in Section \ref{sec:vtx_impinge}.  Consider a vast collection of cases with $G \in [-10,10]$, $d/c \in [0.5,2]$, $h/c \in [-0.5,0.5]$, $\alpha \in \{20^\circ,30^\circ,40^\circ,50^\circ,60^\circ\}$, amounting to a total of 200 cases (40 cases per each angle of attack) with 1200 snapshots for each case.  We perform an autoencoder-based compression of the whole flow field data with the flow field variable $\boldsymbol{q}$ being the vorticity variable $\boldsymbol{\omega}$ and $\boldsymbol{g}$ being the lift $C_L$ experienced by the airfoil.  Hundreds of these cases of time-varying vorticity fields from the extreme vortex-airfoil interactions are found to be captured by a three-dimensional latent variable $\boldsymbol{\xi}$ that resides at the bottleneck of the autoencoder.  This surprising finding reveals that even though there is a large degree of freedom for the overall dataset, there is a three-dimensional representation for the seemingly vast collection of complex vortex dynamics.  The small three-dimensional $\boldsymbol{\xi}$ is found to be sufficient to reproduce all of the flow fields in the training data set \cite{Fukami:NC23}.

\begin{figure}
	\includegraphics[width=0.99\textwidth]{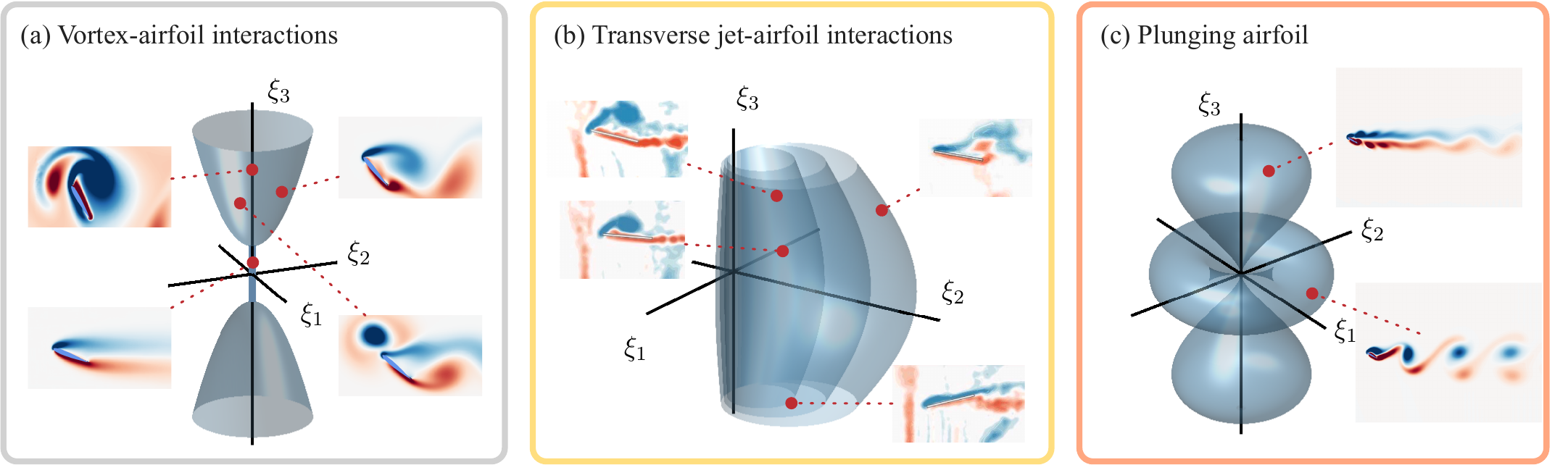}
	\caption{Schematics of low-dimensional extreme aerodynamic attractors revealed from observable augmented convolutional neural network-based autoencoders for (a) vortex-airfoil interactions \cite{Fukami:NC23,Fukami:JFM24}, (b) transverse jet-airfoil interactions \cite{Smith:JFM24}, and (c) plunging airfoil \cite{Odaka:AIAA25}.}
    \label{fig:manifold}
\end{figure}

The latent variable $\boldsymbol{\xi}$ evolves around an attractor illustrated in figure \ref{fig:manifold}(a).  The attractor shows two paraboloids connected by a thin tube, making the overall structure resemble an hourglass.  The attractor for $\alpha > 20^\circ$ exhibits unsteady K\'arm\'an shedding for unperturbed cases.  For $0 \le \alpha \lesssim 20^\circ $, the steady flow (or weakly oscillating flow) collapses to the thin connecting tube.  The dynamics for negative angles of attack are shown by the bottom part of the attractor.  The distribution of latent variable $\boldsymbol{\xi}$ for the disturbed cases collapses around this attractor \cite{Fukami:NC23}.  Observations suggest that this attractor is an inertial manifold, but requires additional discussions.  The coordinate in the radial direction captures the amount of vorticity generated around the airfoil, and the azimuthal direction describes the phase of the dynamics (timing).  Moreover, the axial direction represents the effective angle of attack of the airfoil.

Revealing this attractor is beneficial not only in understanding the low-dimensional nature of extreme aerodynamic response but also in developing modeling and control techniques.  Since the gust disturbance dynamics evolve around the attractor, the dynamics can be projected onto a two-dimensional surface (submanifold \cite{Haller:ND16}), requiring only the radius and phase variables to model extreme vortex-airfoil interactions.  One can also develop a control law using this model to attenuate the transient effects on the airfoil \cite{Fukami:JFM24}.  Examining the physics on this manifold enables us to depart from analysis and control based on time and instead focus on event-based analysis and control.

The attractors can also be found for other types of gust disturbances.  Particle image velocimetry measurements from a tow-tank experiment for a translating wing being hit by a transverse gust \cite{Sedky:PRF23} were analyzed with another autoencoder that incorporated topological data analysis (persistent homology) \cite{Smith:JFM24}.  This analysis revealed another type of attractors shown in figure \ref{fig:manifold}(b) and reformulated transient flows as cyclic phenomena.  Moreover, additional attractors have also been found for different types of airfoil wakes generated by plunging motions, as illustrated in figure \ref{fig:manifold}(c).  The plunging airfoil wakes with an auxiliary variable of drag (or thrust) were compressed to atomic orbit-like data distributions in latent space \cite{Odaka:AIAA25}.  We refer readers to \cite{Smith:JFM24,Odaka:AIAA25} for detailed interpretations of these attractors.

This zoo of attractors provides a fundamental basis for characterizing and understanding key features of extreme aerodynamic flows.  Given these attractors, there is an open question on how these insights can be merged to uncover the landscape of global state representation that connects different types of extreme aerodynamic gusts in general.  
Currently, efforts are underway to find the low-dimensional behavior of three-dimensional turbulent extreme aerodynamic flows.  Preliminary results suggest that there is indeed a low-dimensional description to capture the transient vortex dynamics.  With tip effects in three-dimensional flows, the latent space analysis may require additional dimensions to fully capture the complex dynamics over time.

The autoencoder discussed here was trained with only single-gust cases.  In the vortex-airfoil interaction case, the impact of multiple vortices mimicking a continuous gust was also analyzed with the same autoencoder.  Due to the use of convolutional neural network layers, the autoencoder was able to process the multiple vortices and accurately reconstruct the flow field and the lift trajectory over time \cite{Fukami:NC23}.  While the training dataset did not contain multiple vortices impacting the wing at once but the training data held key information about multi-vortex interactions observed between the gust vortices and the generated vortices.  This type of results confirms the strength of nonlinear machine learning techniques, which would not have been achievable with traditional linear analysis techniques.

\section{Vortex Dynamics}
\label{sec:vortex_dynamics}

In extreme aerodynamics, the gust disturbance holds significant strength, imposing very high levels of unsteadiness on the flying object.  During this encounter, vorticity is generated along the no-slip surface due to the imposed acceleration and tangential pressure gradient \cite{Hornung89,Wu06}.  For incompressible flow, the source of vorticity (wall-normal vorticity flux) $J_n$ at the surface is 
\begin{equation}
    J_n
    = - \hat{\boldsymbol{\tau}} \cdot \frac{d \boldsymbol{V}_\text{wall}}{dt} - \frac{1}{\rho} \left( \frac{\partial p}{\partial \tau} \right)_\text{wall},
    \label{eq:vt_source}
\end{equation}
where $\boldsymbol{V}_\text{wall}$ is the wall velocity, $p$ is the pressure variable, and $\boldsymbol{n}$ and $\boldsymbol{\tau}$ denote the wall-normal and tangential directions, respectively.
Given that the disturbance velocity changes quickly across a short characteristic length scale comparable to that of the body, the acceleration and deceleration that a body experiences within a short time are significant in their magnitudes.  Furthermore,  these disturbances can cause massive flow separation over the wing with a strong pressure gradient along the surface.  Consequently, extreme disturbances induce a high flux of vorticity to be generated along the surface, leading to the creation of large-scale vortical structures.

With strong vortical disturbances generally holding coherent vortex cores \cite{Barnes:AIAAJ18,Barnes:AST18,Biler:AIAAJ21,Fukami:PRF25,Devenport:JFM96}, the vorticity generated during the unsteady process enters the flow in a locally two-dimensional fashion, resulting in a vorticity sheet.  This sheet rolls up into a coherent vortex during the accelerating phase of the unsteady gust encounter.  As flow under acceleration tends to stabilize the flow, the flow structures generated during this phase tend to exhibit laminar features with coherent cores.  In contrast, instabilities can develop greatly during the decelerating phase of the encounter \cite{Linot:JFM24}, breaking up the disturbance and relaxing the vortical structures around the wing.

\begin{figure}
	\centering
	\includegraphics[width=0.9\textwidth]{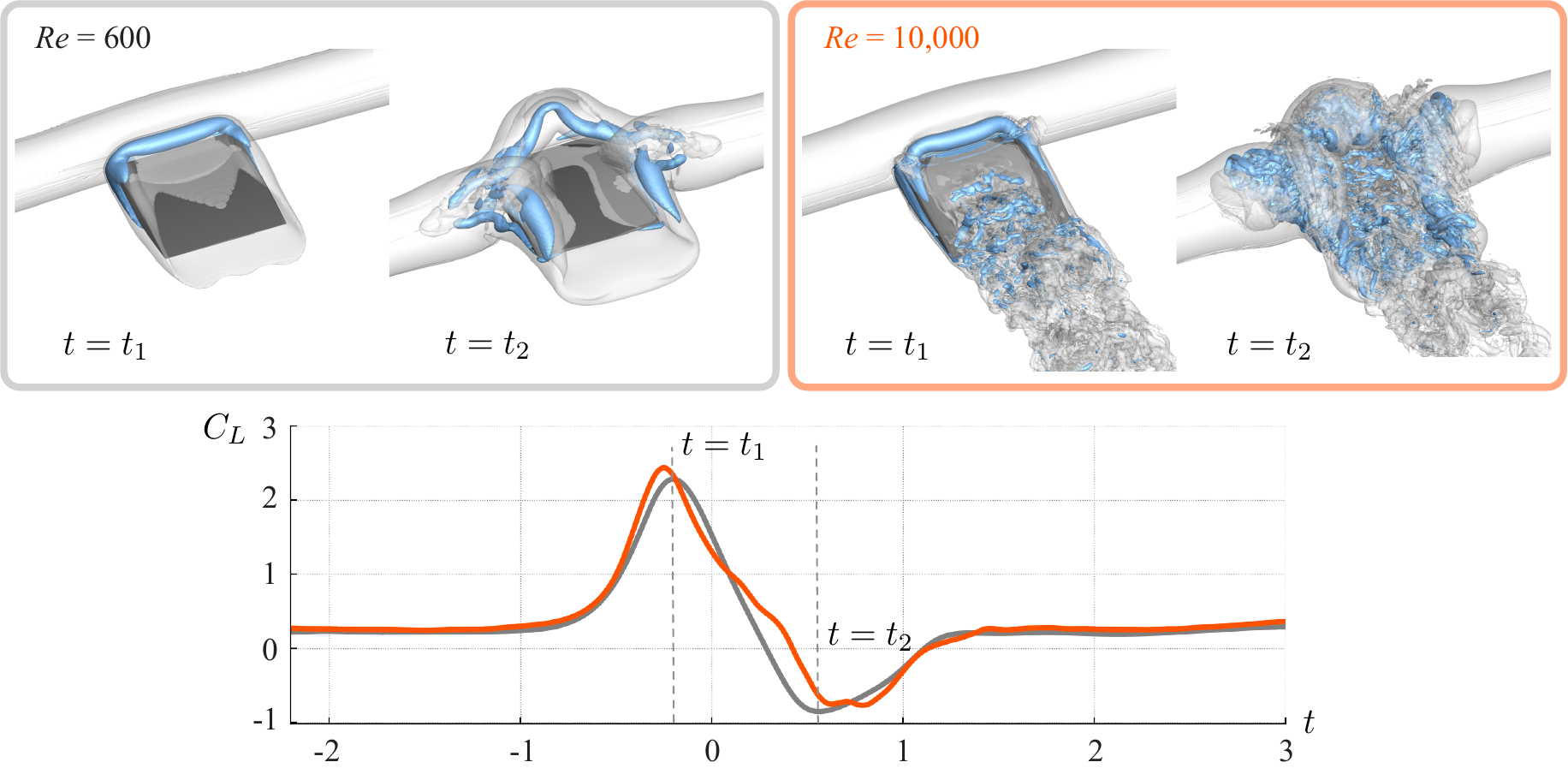}
	\caption{Extreme vortex gust of $(G, d/c, h/c) = (2, 0.5, 0)$ hitting a square NACA 0015 wing at $\alpha = 14^\circ$ for $Re = 600$ and $10,000$ \cite{Odaka:JFMXX2}.  Visualized are the three-dimensional vortical structures with $Q$-criteria ($Q = 100$, blue) and vorticity ($\| \omega \| = 7$, gray) isosurfaces.  Corresponding lift histories are shown at the bottom.}
    \label{fig:AR1}
\end{figure}

During an extreme gust encounter, the vortex core evolution closely follows that of low-Reynolds-number flow due to the cohesive nature of the cores and the short advection time scale ($\sim c/u_G = c/(G U_\infty)$) compared to the longer viscous time scale ($\sim c^2/\nu$).  To demonstrate this point, let us consider the extreme gust vortex-finite wing encounters with $G=2$ at chord-based Reynolds numbers of $Re = 600$ and $10,000$ \cite{Odaka:JFMXX2}, as presented in figure \ref{fig:AR1}.  
Here, we visualize the vortical structures at the time of maximum ($t = t_1$) and minimum ($t=t_2$)  lift.  Despite the Reynolds numbers being quite different in magnitude, there is similarity in the vortical structures.  Most of the difference lies in the presence of small-scale turbulent structures.  However, the vortex cores (blue) remain similar throughout the process.  In fact, the force element analysis \cite{Chang:PRSLA92,Lee:JFM12,Zhang:JFM20,Menon:PRF22} reveals that almost all contributions are from similar coherent vortical structures.  The incoherent turbulent structures do not significantly contribute to the large transient force.  This is also clear from the similarity exhibited by the two lift history curves shown in figure \ref{fig:AR1}.  The striking similarities observed in the extreme gust response across Reynolds numbers tell us that the fundamental understanding of massively separated low-Reynolds-number laminar flow is paramount in deepening our knowledge of extreme aerodynamic flows for higher Reynolds numbers and those with complex problem settings.

In gust encounters, the formation of vortical structures and the emergence of instabilities serve an important role in the overall transient dynamics.  The growth of the leading-edge vortex and tip vortices in particular influences the large lift and drag fluctuations over short time duration \cite{Taira:JFM09,Lee:JFM12,Odaka:JFMXX}.  To aid the understanding of such vortex evolution processes, stability and modal analysis techniques \cite{Schmid01,Taira_etal:AIAAJ17,Colonius25} can be beneficial in revealing the source of unsteadiness.  However, these techniques are generally developed for time-invariant base flows, and their applications become limited for time-varying base flows beyond time-periodic flows \cite{kutz2016multiresolution,Schmidt:JFM19,Souza:TCFD24,Kai:arXiv25,Babaee:PRSA16,Zhong:JFM25}.  The study of transient flows with time-varying base flows calls for the development of novel theoretical and data-driven techniques \cite{Linot:FDR25}.

Traditional aerodynamics with small $G$ considers gust perturbations dynamics centered around the wing aerodynamics being the dominant physics.  However, extreme aerodynamics is defined for flows with gust ratios $G > 1$, in which case the disturbances can be considered as the primary feature and the airfoil being the perturbation to the gust.  This change in perspective may pose new problem settings that may benefit from theoretical vortex dynamics \cite{Saffman92,Eldredge19,Baddoo:JFM21} and can seek inspirations from analysis techniques used in atmospheric and oceanographic fluid dynamics that examine vortex dynamics across a very large range of spatiotemporal scales.

\section{Sensing and Estimation}

Gaining situational awareness in an extreme flight environment is critical for an air vehicle to stay aloft.  Approaching extreme disturbances must be detected as soon as possible to initiate evading maneuvers or activate flow control to mitigate large-amplitude transient effects on the aircraft.  Taking quick and effective action against extreme aerodynamic disturbances is crucial.  One approach is to use an ensemble Kalman filter with low-order point vortex models to estimate the state of the disturbed flow around an airfoil \cite{LeProvost:PRF21}.  The use of criteria such as the leading-edge suction parameter \cite{Ramesh:JFM14} could support the vortex models to capture leading-edge separation observed in extreme aerodynamics with some care towards strong unsteadiness \cite{Deparday:PF19}.

If a sufficient amount of training data is available, machine learning can be used to estimate the flow field around the wing based on sparse sensor measurements.  For instance, a limited number of pressure sensors on a wing surface can be used to estimate the instantaneous vorticity field around the wing as well as the aerodynamic forces \cite{Zhong:TCFD23}.  For flow field reconstruction, it is possible to directly relate the pressure measurements to the flow field through a neural network.  From this state estimation analysis, an approaching vortical gust can be detected at least a couple of chord lengths upstream of the airfoils.  Moreover, force estimation has also been performed for experimental measurements at modest Reynolds numbers.  A tow-tank experiment demonstrated that it is possible to develop a machine learning based model that can estimate forces acting on a delta wing during unsteady maneuvers that model transverse and lateral gust encounters even in the presence of turbulence and noise \cite{Chen:AIAAJ24}.  The derived machine learning model can also be used to identify sensors that are most influential in predicting the force.  

The above approaches directly relate surface measurement signals to the flow field or forces on the body.  Another possibility is to take the sensor measurements and relate them to the low-dimensional latent space representation of extreme aerodynamic flows (recall figure \ref{fig:manifold}).  In this case, only an encoder from the sensors to the latent variables $\boldsymbol{\xi}$ needs to be trained if the decoder is available from prior training \cite{Odaka:AIAA25}.  Components of machine learning models can be used effectively in these cases without having to retrain the entire network.  Also beneficial is the quantification of uncertainty in machine learned models for extreme aerodynamic flows \cite{Mousavi:JFM25,Maulik:PRF20}.  With this type of quantification, the effects of noise and model uncertainty can be assessed, which can support flight trajectory planning and flow control designs.

Thus far, we have discussed sensing and estimation of extreme aerodynamic flows around a wing.  However, we can consider gaining situational awareness at a larger scale.  There are now efforts to reconstruct the local flow in urban canyons and other regions alike \cite{Yasuda:BE23,Yasuda:UC25,Jaroslawski:IJHFF25}.  As microweather can generate strong local fluctuations around these complex terrains, these flow estimations from either coarse weather measurements or sensors can be of benefit to air vehicles navigating in such airspace.  Sensors on building walls or traffic light poles may effectively serve as input to detect adverse flow structures in the flight path of small-scale air vehicles.  Machine learning based flow estimation and super-resolution analysis \cite{Fukami:TCFD23,Yasuda:BE23,Yasuda:UC25} can be especially effective in unveiling the complex turbulent flow fields.  There also exists a flow reconstruction technique that can utilize moving sensors and changing sensor populations that can adapt to dynamically changing situations \cite{Fukami:NMI21}.  With the merging of onboard sensors from the aircraft and those from the ground or nearby aircraft, a greater situational awareness can provide advanced warning to the flying vehicle to better manage the approaching extreme gusts.

\section{Flow/Flight Control and Trajectory Planning}

Next, let us discuss some approaches to managing the extreme aerodynamic effects on an aircraft.  Broadly speaking, there are three main approaches: (1) flow control to modify the behavior of the flow, (2) flight control to stabilize flight, and (3) trajectory planning to safely and efficiently travel from the origin to the intended destination.

Active flow control \cite{Joslin09,GadElHak00} can reduce the large transient forces imposed by the extreme aerodynamic disturbances.  However, we should remind ourselves that the amplitude of such transient forces can be of enormous magnitude, fluctuating over a very short time scale. Because active flow control aims to use minimal input to achieve maximum changes, the approach generally takes advantage of inherent flow instabilities to amplify input actuation.  This input amplification takes place over the instability timescale. In an extreme aerodynamic situation, it would be necessary to mitigate an extremely large transient effect quickly within approximately two convective time units. 

Presently, the development and implementation of energy-efficient or effective fast-acting actuation \cite{Cattafesta:ARFM11} strategies remains an open question. For an actuator to act on the flow quickly, it will likely need to quickly modify the flow, which will likely require a large energy input.  One possibility is to swiftly change the circulation of the wing by using flaps as actuators \cite{Medina:AIAAJ17}.  While the energy requirement to move flaps quickly may be large, we do have a much greater willingness to pay for operating in extreme airspace.  

There are some promising approaches.  Recently, active flow control was studied for a two-dimensional gust-airfoil interaction problem discussed in Section \ref{sec:vtx_impinge}.  A low-dimensional model, developed based on phase-amplitude modeling, served as a basis for a fast-acting jet-based control, which was able to attenuate the transient effects over approximately two convective times with a reduction of about $30\%$ in extreme lift increase \cite{Fukami:JFM24}.  The quick-acting nature of the control was achieved by using concepts from phase synchronization.  While $30\%$ reduction may seem small at first glance, this level of attenuation is substantial because extreme aerodynamics deals with a very high level of disturbances.   Ongoing efforts that examine the receptivity of gust vortices and their relationship to separation and lift generation mechanisms may also support the flow control endeavor.

An alternative approach to mitigate the effect of gust is to change the vehicle attitude.  In experimental studies of transverse jet hitting a translating airfoil, dynamically changing the angle of attack is shown to significantly reduce lift fluctuation experienced by the airfoil for $G \le 0.71$ \cite{Xu:AIAAJ23,Sedky:PRF23}.  The use of Wagner-K\"ussner framework for a closed-loop proportional control is shown to minimize the transient lift for cases with the input of the second derivative of the angle of attack \cite{Sedky:PRF23}.  Similar maneuvers with dynamically changing the angle of attack were also examined using reinforcement learning \cite{Liu:AIAAJ25} with observed reduction in lift fluctuation.  The combination of flow control and flight control holds the potential to enable sufficient suppression of extreme gust effects at even high gust ratios.  While the use of deep learning-based flow control efforts has not been prevalent for extreme aerodynamics, the foundations to do so appear to be ripe \cite{Rabault:JFM19,Vinuesa:Fluids22,Linot23,Deda:PRF24}.

Trajectory planning is also another important aspect of ensuring sustained flight in the violent extreme airspace.  Traditional flight plans from the origin to the destination generally involve smooth, almost linear trajectories.  For smaller vehicles navigating in extreme aerodynamic environments, frequent avoidance of dangerous flow structures will likely be necessary.  However, in some cases, flying into or in the vicinity of strong disturbances may be beneficial if the vortical forces exerted from the disturbances can be used to the advantage of the vehicle.  

It has been shown that vortical structures in canonical wakes can be used beneficially to navigate with reduced power requirements.  Navigating upstream in a two-dimensional wake can be helpful when crossing from one side of the wake to another \cite{Gunnarson:NC21}.  In a three-dimensional circular cylinder wake, another study further revealed that the streamwise vortical structures can be further taken advantage of by performing a corkscrew-type maneuver to further enhance the efficiency of navigating across a wake \cite{Godavarthi:Flow25}. In this study, model predictive control with a relatively short time horizon was used to determine the optimal vehicle trajectory.  This provides promise in incorporating sensor-based state prediction, which can be performed with a very small number of sensors \cite{Zhong:TCFD23,Chen:AIAAJ24}.  In a more applied setting, reinforcement learning based trajectory planning has also been demonstrated to find a flight path around a tandem building in a turbulent environment \cite{Tonti:JCP25}.

\section{Aerodynamic design}

Traditional wings have been designed with steady (or quasisteady) flight in mind \cite{WingTheory}.  These wings have evolved to find the geometry that achieves a high lift-to-drag ratio in pursuit of efficiency.  However, aircraft that are required to fly in extremely gusty environments may not benefit from traditional wing designs since their objective is not necessarily to fly efficiently but to navigate through a highly unsteady airspace while being impacted by strong vortices and jets.  This means that the wing geometry suitable for extreme aerodynamic flight environments would likely be different from standard wing designs.  In fact, aircraft and wing design theory and guidelines are essentially nonexistent for aircraft to fly in highly unsteady flows.  Hence, there is an open call for efforts to develop air vehicle designs that can maneuver under strong atmospheric disturbances.

\begin{figure}
	\centering
	\includegraphics[width=0.95\textwidth]{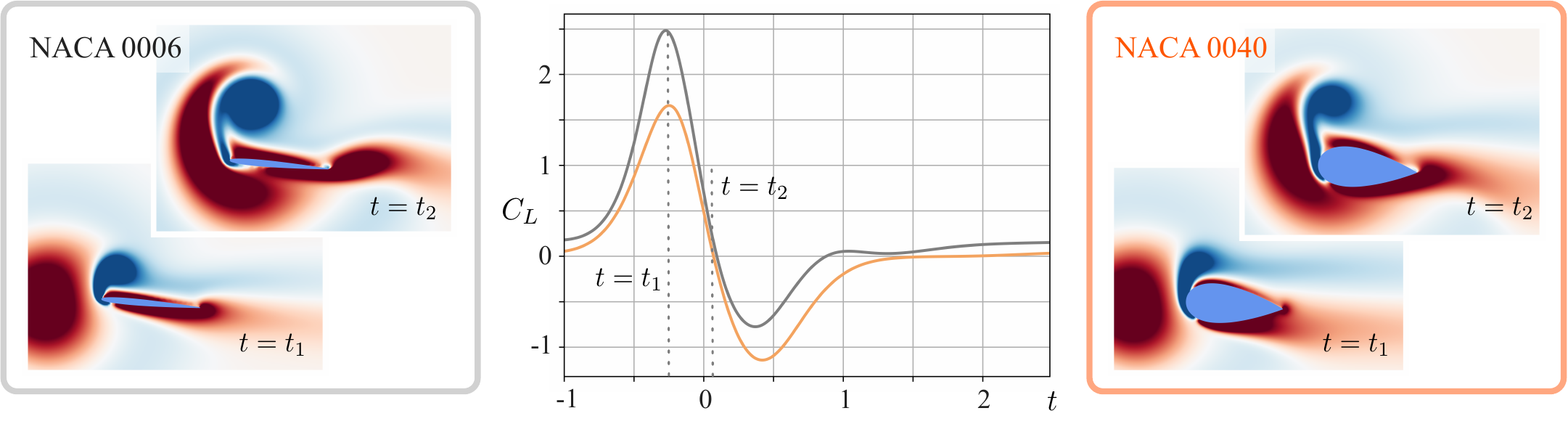}
	\caption{Extreme gust vortex responses by a thin (NACA 0006) and thick (NACA 0040) airfoils \cite{Lopez-Doriga:PRFXX}.  Vorticity fields are shown shortly after impact by a vortex gust of $(G,d/c,h/c,\alpha) = (2,1,-0.1,0^\circ)$.  Corresponding lift histories are also shown.}
    \label{fig:thickness}
\end{figure}

Let us show one of the basic effects of the airfoil geometry on extreme aerodynamic gust response.  We consider the role of thickness has on transient lift increase during a strong vortex impact ($G=2$ and $Re = 100$) \cite{Lopez-Doriga:PRFXX}.  Shown in figure \ref{fig:thickness} are the differences in the vorticity fields for this vortex impacting a thin NACA 0006 airfoil and a thick NACA 0040 airfoil.  After the vortex impacts the thin airfoil, we observe a strong leading-edge vortex (blue) generated.  This large vortex imposes a very large lift increase within one convective time.  In contrast, a weaker leading-edge vortex is produced by a thick airfoil.  The differences in the strengths of the generated leading-edge vortices for the two airfoils become evident once the leading-edge vortices start to separate from the airfoils.  

The thicker airfoil experiences reduced lift increase from the relatively weaker leading-edge vortex.  The reason for this significant difference in the vorticity fields and the associated lift behavior comes from the curvature of the leading edge.  As discussed in Section \ref{sec:vortex_dynamics}, the vorticity flux entering the flow field from the no-slip surface is dependent on the acceleration of the body and the pressure gradient along the surface (see equation \ref{eq:vt_source}).  With the curvature being lower for the thick airfoil, a reduced level of vorticity is generated around the leading edge, which leads to the creation of a weaker leading-edge vortex and hence a lower increase in lift after the gust vortex impact.

We are not necessarily suggesting the selection of thick airfoils (such as those used on wind turbines \cite{WTAcatalogue}) for aircraft flying in extreme gust environments.  What we are aiming to relay here is that the modification of flow physics near the leading edge is important in reducing the lift increase.  
While not shown here, three-dimensional vortical structures can also help in reducing the large transient forces during gust encounters \cite{Odaka:JFMXX}.  Tip vortices can also reduce force fluctuations during gust encounters.  
These observations may lead to concepts that actively or passively alter the leading-edge and tip geometry and, accordingly, the surface vorticity dynamics.  Even though such approaches might have been difficult in the past, there are now possibilities through a novel class of materials and structures \cite{Ge:APL13,Shan:AM15,Treml:PNAS18}.  

Let us also note that once a significant number of extreme aerodynamic flow responses are collected for a range of airfoil and wing designs, machine learning based analysis can be used to find the low-dimensional trends.  In fact, the observable augmented autoencoders (figure \ref{fig:AE}) can also be used for design optimization.  Geometric information can be used as input and outputs to an autoencoder with an appropriate auxiliary variable to identify a low-dimensional latent space for aerodynamic design analysis \cite{Tran:CE24}.  This approach was used to aerodynamically optimize the design of commercial automobiles.  Similar efforts can be taken to analyze and optimize airfoil and planform geometry as well as material properties for air vehicles in extreme aerodynamic environments.  With machine learning based design optimization being a very active area of research, we can expect growth in analogous research activities across engineering \cite{Tangsali:arXiv25}.

Machine learning based analysis can also reveal where training data is scarce and suggest new test designs to expand the data coverage and improve data density.  Establishing a reliable training dataset requires a large number of high-fidelity simulations, which can be computationally demanding.  However, low-fidelity simulations can also supplement the training dataset through multi-fidelity analysis techniques \cite{Peherstorfer:SIAMRev18,Pinti:AIAAJ22}.  
Moreover, given the advancement in automation and 3D printing technology, there are now possibilities for experimental measurements to be collected in mass as well \cite{Mulleners:PRF24}.

At a more applied level, characterization of gusts in the expected flight environments is necessary to perform design optimization for extreme gust-resilient vehicle designs.  With statistical characterization of the gusts, designs should be able to weigh the level of resilience for the different types of extreme gusts.  Uncertainty quantification \cite{Smith14} should also be incorporated into this process to provide bounds on uncertainty for the selected wing design and gust models.  These assessments will enable the estimation of safety margins for structural and flight control considerations. 

We should remind ourselves that extreme gusts are violent in nature, and their effects may not be completely suppressed with a single solution.  However, their transient effects may be reduced through a collective effort by addressing the surface flow dynamics, active flow control, and flight control/planning.  Again, the main objective for flight in extreme aerodynamic conditions is not efficiency but the sustainment of flight.  The resulting flight may be very different from the steady, calm flight that we are used to.  As such, the vehicle may be of a design that is completely different from traditional aircraft.  The new design process will need to fold in the insights we learn from extreme aerodynamics to combat the violent gust disturbances.  Preventing drink spillage or passenger comfort is likely not a top priority for vehicles that operate in extreme aerodynamic conditions.

\section{Outlook}

The necessity to operate small-scale aircraft in highly gusty environments during adverse weather calls for innovations in analysis, modeling, estimation, control, and design techniques for extreme aerodynamics.  Given the strong nonlinearity and the vast amount of parameter space associated with extreme aerodynamics, modern machine learning and data science techniques can support deepening our understanding of physics.  The use of these methods can uncover the underlying flow responses in low-dimensional latent space, which suggests the possibility of developing real-time flow estimation, modeling, and control of air vehicles in the violent air space.  Before closing this paper, let us comment on a few additional topics that can take advantage of data-driven approaches to make headway.

One area of research we have not discussed yet is the development of aerodynamic models.  There have been numerous semi-analytic models derived over the years to predict aerodynamic effects on wings against perturbations \cite{wagner1924entstehung,glauert1930force,theodorsen1935general,Sears:JAS41,Hoblit88,Katz01,Leishman06,Lucia:PAS04}.  They perform well for low to modest gust ratios in many cases.  However, their uses need to be carefully considered for extreme aerodynamic conditions, as they were not originally designed for large gust ratios.  With extreme aerodynamic data becoming available, regressions can be performed to tune the model parameters.  However, modifications to the semi-analytic models will likely be needed by considering additional terms \cite{Brunton:PNAS16} or by adopting deep learning techniques \cite{Bishop24,Racca:JFM23}.  Alternatively, models can be established in low-dimensional latent space (Section \ref{sec:manifolds}) by leveraging neural network models that capture various types of nonlinear dynamics \cite{Linot:Chaos22,Fukami:JFM24,Smith:JFM24,Yawata:PREXX}.  Moreover, there are approaches that can patch different physical regimes in hopes of establishing a model that covers a wide range of operating conditions \cite{lozano2023machine,oulghelou2024machine}.  Extreme aerodynamic model development will be a critical area of research, as we have seen for turbulence modeling \cite{duraisamy2019turbulence,Nair:PRF25}.  

Biological flyers provide us with inspiration from how they fly in gusty conditions \cite{Treidel:ICB24}.   Because these flyers do not fly in adverse conditions unless provoked or trained \cite{Ortega:JEB13}, collecting measurements and studying their behavior can be challenging.  However, there are recent studies on birds and insects managing large transient forces during flight \cite{Bamford:BB24,Zhang:AIAA17}.  By learning how these creatures quickly adjust themselves to adapt to the surrounding atmospheric disturbances can help develop controllers.  
Combining the gained insights from bioflyers, developments in extreme aerodynamics, and advances in novel materials and structures may inspire innovative vehicle designs.

This paper mostly considered the gust ratio as a primary parameter to capture the effect of extreme gusts.  Exploring additional non-dimensional parameters that best capture the gust effects on the forces and the flow fields is warranted.  In addition to the theoretical Buckingham Pi analysis \cite{Stutz:AST23}, there is an opportunity for data-driven analysis to uncover insightful non-dimensional parameters \cite{Bakarji:NCS22} and nonlinear relationships among them to deepen our understanding of the highly unsteady flow behavior \cite{Fukami:JFM24b}.  By identifying physically important non-dimensional parameters, there is a chance for extreme aerodynamic physics to be expressed in a universal manner to advance the overall field of research.

Another area of extreme aerodynamics that has received limited attention is the influence of three-dimensional effects on gust-wing interactions \cite{Garmann:JFM15,Barnes:AIAA19,Qian:EF22,Odaka:JFMXX,Odaka:JFMXX2}.  It is known that wing planform parameters, including the aspect ratio, sweep, taper, and twist, can influence the three-dimensional wake dynamics \cite{Taira:JFM09,Zhang:JFM20a,Ribeiro:JFM23,Pandi:JFM25}.  Examining the influence of strong gust disturbance on such three-dimensional wakes is computationally taxing and will require strategic planning of high-fidelity numerical simulations and experimental measurements to capture as much of the rich dynamics as possible.  Smart sampling techniques may help plan for these parametric studies to establish a dataset that can be used to extract the underlying three-dimensional extreme aerodynamic behavior for a vast group of wings.

Flying in an extremely gusty environment requires significant energy for propulsion and control.  As such, innovative techniques that can possibly attenuate gust effects with active input are critically needed.  Taking advantage of aeroelasticity may prove useful in helping vehicles achieve a larger flight envelope and energy savings \cite{Fernandez:JFS22,Thompson:AIAAJ25}.  The structural dynamics for flexible wings and their interactions with the surrounding fluids will likely be highly complex, with potential for data-driven techniques to greatly support the analysis \cite{Fonzi:PRSA20,Nair:TCFD23,Hickner:AIAAJ23}.

There are additional opportunities for machine learning based techniques to play an important role.  Flow phenomena that are difficult to describe with governing equations are prime problem settings for which data-driven techniques can support their characterization and modeling.  As extreme aerodynamic flight conditions occur during adverse weather at very low altitude, rain, snow, and ice can introduce additional complications for sustained flight, especially with ice accretion on lifting surfaces \cite{Thompson:JA96,Bragg:PAS05,Cao:PAS14,Cao:AST18,Liu:AST18,Dhulipalla:AIAA24}.  The resulting complex multi-phase flow, which is not easy to describe with governing equations, can grossly modify the wing geometry and its aerodynamic characteristics.  Machine learning techniques may reveal how the multi-phase flow can alter the extreme flow dynamics and offer clues on how to keep the vehicle minimally affected by rain, ice, or dirt.

With the operating condition being vastly different from large-scale aircraft, vehicle designs for extreme aerodynamic flight environments may evolve differently.  The operation of such an aircraft will likely be unconventional, as cruise-like flight may not be achievable, and the objective may consist of staying in the air no matter the cost or the level of comfort.  We will also need to take full advantage of the capabilities of all subsystems to make flight possible.  This means gust detection, flow/vehicle control, trajectory planning, and vehicle design need to be performed in an orchestrated and innovative manner.  There may even be possibilities to consider information sharing from surrounding sensors and other air vehicles to gain global state awareness.  Faced with the technical challenges of extreme aerodynamic flows, unconventional data-driven approaches are likely required to further understand extreme aerodynamic flows and translate the insights into making flight possible in the violent airspace that was traditionally avoided.  We hope this paper stimulates discussions on extreme aerodynamics and provides ideas to enhance the safety of air vehicles that must fly in adverse conditions.

\section*{Acknowledgments}

This work benefited from extensive discussions with Oksan Cetiner, Jeff~D. Eldredge, Kai Fukami, Justin Jaworski, Anya R.~M.~Jones, Barbara Lopez-Doriga, Hiroto Odaka, and Douglas R.~Smith, as well as the members of the NATO AVT 347 and 426 groups.  The term ``extreme aerodynamics" was coined by D.~R.~Smith to describe the violent nature of gust-wing interactions at high gust ratios.  The author gratefully acknowledges the support provided by the U.S.~Department of Defense Vannevar Bush Faculty Fellowship (N00014-22-1-2798) and the U.S.~Air Force Office of Scientific Research (FA9550-21-1-0178).  This article is based partly on the presentation given by the author at the 26th International Conference of the Theoretical and Applied Mechanics (ICTAM 2024).

\bibliographystyle{unsrt-init}  
\bibliography{Taira_refs,Taira_refs2}

\end{document}